\documentstyle[prd,aps]{revtex}
\title{\bf A Path Integration Approach to Relativistic
           Finite Density Problems \\and Its Particle Content}
\author{S. Ying}
\address{Physics Department, Fudan University\\
Shanghai 200433, China}
\date{\today}

\narrowtext
\draft

\begin{document}
\maketitle
\begin{abstract}
  A path integration formulation for the finite density and
  temperature problems is shown to be consistent with the
  thermodynamics using an 8
  component ``real'' representation for the fermion fields
  by applying it to a free fermion system. A relativistic quantum
  field theory is shown to be smoothly approached at zero temperature
  by a real-time thermal field theory so derived even at a finite
  density. The analysis leads to a 
  new representation for the fermion fields which is shown to be 
  inequivalent to the conventional 4 component
  theory at the quantum level by having a mirror universe
  with observable effects and to be better behaved at short distances.
\end{abstract}

The relativistic quantum field theory (QFT) at finite density and
temperature is currently a viable tool to study the real-time processes
of a relativistic system under extremely conditions like in heavy ion
collisions, in astronomical and cosmological processes, etc. A
sufficiently general formulation of the relativistic QFT at
conditions different from zero temperature and density is necessary
so that it can handle the cases of non-perturbative
spontaneous symmetry breaking and particle production and is consistent
with thermodynamics in real-time. Such an extension is not trivial.

One of the non-perturbative treatments of the relativistic QFT is based upon
Feynman--Matthews--Salam (FMS) path integration \cite{FMS}. The FMS formalism
expresses time evolution between initial and final states in terms
of an integration over paths that connect the initial and final states
with a weight determined by the action of each particular path.
The results of the path integration at a formal symbolic level are
not uniquely defined in the Minkowski space-time
due to the presence of singularities. Their uniqueness is
determined by the fact that the
derivation of the FMS path integration representation of the
evolution operator implies a particular ordering of the intermediate
states which defines how the singularities of the formal results are
going to be handled. Thus the FMS path integration formalism is defined
by not only the formal expressions that contain singularities but
also by the ``causal structure'' following the time ordering of
the physical intermediate states.

 One form of the thermal field theory (TFT)
was developed in Ref. \cite{Matsu} in the Euclidean space-time
with the time variable playing the role of the inverse temperature. Such
a formalism, which is based on the periodic (antiperiodic)
boundary condition for bosons (fermions) fields,
provides the correct thermodynamics for a noninteracting system.
The periodic (antiperiodic) boundary condition for the field
operators is explicitly written as
\begin{eqnarray}
 \widehat\psi(t-i\beta) & = &\eta
e^{-\beta\mu}\widehat\psi(t),\label{Fermion_BC}
\end{eqnarray}
where $\beta$ is the inverse temperature, $\eta=\pm 1$ for bosons and
fermions respectively and $\mu$ is the chemical
potential.  The physical responses of the system to external
stimulations that happen in real time can be obtained by a proper
analytical continuation. Another formulation of the same
problem based upon Eq.
\ref{Fermion_BC} can be derived by a distortion of the Matsubara time
contour, which goes straightly from $0$ to $-i\beta$, to a contour
that contains the real time axis extending from negative infinity to
positive infinity and returning contour below the real time axis
somewhere between $0$ and $-i\beta$ that parallels the real time axis
(see, e.g., Ref. \cite{LandsWeer}).

Consistency requires that the results of
the TFT go smoothly to that of a relativistic QFT in real time
at zero temperature (or $\beta\to\infty$). The question is whether or
not it actually happens, especially for fermions. To investigate such a
question, let us study the thermodynamics of a free
fermion system at finite density and zero temperature using relativistic QFT
defined by a FMS path integration formalism.

The thermodynamics of a system with variable fermion number density
is determined by the grand potential
$\Omega$ defined in the following
\begin{eqnarray}
  e^{-\beta \Omega} &=& < e^{-\beta(\widehat H_0 - \mu \widehat N)}>,
\label{PartFun1}
\end{eqnarray}
where $\widehat N$ is the fermion number density operator and $\widehat H_0$
is the (free) Hamiltonian of the system. The result for $\Omega$ is
well known from elementary statistical mechanics; it is $U-TS-\mu N$
with $U$ the internal energy, $T$ the temperature, $S$ the entropy and
$N$ the average particle number of the system.

In the quantum field theoretical investigation of the same system at zero
temperature, there is
another representation for the right hand side (r.h.s.) of Eq.
\ref{PartFun1} in terms of
Feynman--Matthews--Salam \cite{FMS} path integration
\begin{eqnarray}
  \lim_{\beta\to\infty} <e^{-\beta \widehat K}> = const\times \int
  [D\psi][D\bar\psi] e^{-S_E},
\label{PartFun2}
\end{eqnarray}
where $\widehat K\equiv \widehat H_0 - \mu \widehat N$, ``$const$'' is so
chosen
that $\Omega=0$ at zero fermion density. $S_E$ is the Euclidean action
of the system; extending the rules of Ref. \cite{Mehta}
to finite density cases, it can be
obtained from the Minkowski action for the system by a continuous
change of the metric. The result is
\begin{eqnarray}
  S_E = \int d^4 x \bar\psi \left ( i\rlap\slash\partial + \mu
  \gamma^5 - m \right ) \psi,
\label{EuclAct1}
\end{eqnarray}
where $x^\mu$ is the (4-dim) Euclidean space-time coordinates,
$\gamma_E^0=i\gamma^5$ and $m$ is the mass of the fermion. $\Omega$ is
obtained by identifying $\beta$ with $x_4$ (the Euclidean time) that
confines the system in the temporal direction before the
thermodynamic limit.  From Eq. \ref{EuclAct1}, the path integration
can be carried out immediately. The result \cite{Kapusta} is
\begin{eqnarray}
  \Omega' &=& -\lim_{\beta\to\infty} {1\over \beta} \mbox{Ln}
  \mbox{Det} \gamma_E^0\left (i\rlap\slash\partial + \mu \gamma^5 -
  m \right ) - \ln(const) \nonumber \\ &=& - \lim_{V\to\infty} 2 V
  \int_0^\infty {d\mbox{\boldmath $p$} \over 2\pi^2}
   \mbox{\boldmath $p$}^2\int_{-\infty}^{+\infty} {dp_4\over
   2 \pi}
   \ln {p_+^2 + m^2\over p^2+m^2}
\label{EuclAct2}
\end{eqnarray}
with the Euclidean 4 momenta $p^\mu=(p_4,\mbox{\boldmath $p$})$ and
$p_+^\mu = (p_4-i\mu,\mbox{\boldmath $p$})$. The result for $\Omega'$
is finite despite the infinite integration in momentum. For example, in
case of zero mass
$PV = -\Omega'  = V  \mu^4/12\pi^2$,
which is expected from the elementary statistical mechanics.

The Euclidean energy
$p_4$ can be treated as a complex number so that the $p_4$ integration
in the complex Minkowski energy $p_0$ plane is represented by contour
$C_E$ in Fig. \ref{Fig:Cont1}.  In the study of the real-time
processes of the system, Minkowski action should be used. The
original $p_0$ integration contour $C_{FMS}$ corresponding to the FMS
path integration in the Minkowski space-time is also shown in Fig.
\ref{Fig:Cont1}.  There are also a pair of contours $C_+$ and $C_-$
shown in Fig. \ref{Fig:Matsu},
called the quasiparticle contour, that contribute to the real-time
response of the system in the $\beta\to\infty$ limit.
Contours $C_{FMS}$, $C_E$, $C_+$ and $C_-$
in Fig. \ref{Fig:Matsu} belong to the
same topological
class of contours having the same FMS causal structure.
Consistency requires the equivalence of the set of contours
$C_{FMS}$, $C_E$, $C_+$ and $C_-$ for the physical quantities.
Eq. \ref{EuclAct2} unfortunately fails to meet this requirement due to the
fact that the imaginary part of the logarithmic function falls
off as $O(\mu/|p_0|)$ on the physical $p_0$ sheet (the imaginary
part of the logarithmic function on the edges of its branch cuts on
the physical sheet is shown in Fig. \ref{Fig:Cont1}). This causes the
results obtained from doing the $p_0$ integration on
the above mentioned set of contours different from each other since
the integration on the sections of the large circle of contours
that connecting them has a non-vanishing value.

Explicit computation of Eq. \ref{EuclAct2} on contour $C_+$ where
only the imaginary part of the integrand with value $\pm i\pi$
contributes (see
 Fig. \ref{Fig:Cont1}), reduces to a form
differs quite drastically from the expected grand
potential of thermodynamics, which is $\Omega=
U - \mu N$. In fact, it diverges.

The cause of the
above mentioned problems is found to be
related to the asymmetric nature of the 4 component representation of
the fermion fields with respect to particles and antiparticles.
An 8 component ``real'' representation for the fermion fields
can be used \cite{Paps}. For a spin $1/2$ particle
with one flavor, its 8 component spinor $\Psi$ can be written as
$\Psi=(\psi_1^T,\psi_2^T)^T$
with the reality condition given by
\begin{eqnarray}
  \Psi = {\cal M}\bar\Psi^T,
\label{realcond}
\end{eqnarray}
where superscript ``T'' represents the transpose. Here ${\cal M}=
O_1\otimes C$
with $C$ the ordinary charge conjugation operator and $O_1$ the first
of three
Pauli matrices $O_{1,2,3}$ acting on the upper ($\psi_1$) and lower
($\psi_2$) 4 components of $\Psi$. In the Minkowski space-time,
$\Psi$ transforms in the same way as $\psi_1$ or $\psi_2$ under the
Poincar\'{e} group, which commutes with $O_i$ ($i=1,2,3$).
Covariant operators can thus be constructed in the same way as in the
4 dimensional representation. For example, the bilinear terms in
$\Psi$ can be written as $\bar\Psi \Gamma\otimes O_\lambda \Psi$,
where $\Gamma$ is any covariant matrix in the 4 dimensional Dirac
space, $\lambda=\{0,1,2,3\}$ and $O_0 = 1$. Only the matrices
satisfying the antisymmetric condition $({\cal M}\Gamma\otimes
O_\lambda)^T= - {\cal M}\Gamma\otimes O_\lambda$ can be selected since
$\Psi$ is a collection of anticommuting Grassmanian numbers in the
path integration language.

In terms of $\Psi$, the invariant action $S_E$
(Eq. \ref{EuclAct1}) for a free massive fermion system
in the Euclidean space is
$
  S_E = {1\over 2}\int d^4 x \bar\Psi \left ( i\rlap\slash\partial +
  \mu \gamma^5 O_3 - m \right ) \Psi
$
and Eq. \ref{PartFun2} becomes
$
  \lim_{\beta\to\infty} <e^{-\beta \widehat K}> = const\times \int [D\Psi]
  e^{-S_E}
$
where the $\bar\Psi$ degrees of freedom are not
functionally integrated due to the reality condition Eq.
\ref{realcond}.  The path integration can again be easily evaluated:
\begin{eqnarray}
  \Omega &=& - \lim_{V\to\infty} V
  \int {d^4 p\over (2\pi)^4} \ln {(p_+^2 + m^2)(p_-^2+m^2)\over
    (p^2+m^2)^2}
\label{EuclAct4}
\end{eqnarray}
with the same order of integration as Eq. \ref{EuclAct2}.
Here $p_-^\mu=(p_4+i\mu,\mbox{\boldmath $p$})$.  Eq. \ref{EuclAct4}
and Eq. \ref{EuclAct2} have an identical value on the contour $C_E$.
They differ on contours $C_\pm$ because the large $p_0$ behavior of the
imaginary part of the logarithmic function in Eq. \ref{EuclAct4} is of
order $O(\mu^2/|p_0|^2)$ on the physical sheet,
which allows the equivalence between the set
of contours $C_{FMS}$, $C_E$, $C_+$ and $C_-$.
By following the
$C_+$ contour shown in Fig. \ref{Fig:Cont1}, it is simple to show that the
resulting r.h.s. of Eq. \ref{EuclAct4} is finite and unique, namely,
$U - \mu N$. It is the zero temperature
grand potential for a free fermion system at density $\bar n= N/V$
expected from thermodynamics.

The correct grand potential for a free system in real-time
can be obtained using
a 4 component representation for the fermion fields in the TFT.
The standard method do a Matsubara summation over discrete (imaginary)
energies implied by Eq. \ref{Fermion_BC}
for the 4 component Dirac spinor $\psi$. Such a
boundary condition for $\psi$ results, however, in different analytic structure
for the path integration than that for the FMS path integration in the real
time.
The grand potential in the Matsubara formalism is
\begin{eqnarray}
  \Omega &=& -\lim_{V\to\infty} {2V\over\beta}
          \int {d^3p\over (2\pi)^3} \nonumber\\
 &&\phantom{-}\, \sum_{n=-\infty}^{\infty}\ln
  {m^2+\mbox{\boldmath{$p$}}^2-\left [ i(2n+1)\pi\beta^{-1}+\mu \right
    ]^2\over m^2+\mbox{\boldmath{$p$}}^2-\left [ i(2n+1)\pi\beta^{-1}
  \right ]^2}.
\label{MatsuAct}
\end{eqnarray}
The sum over Matsubara frequencies can be evaluated by using a contour
integration. For this case, it is
$
  2\pi i\beta^{-1}\sum_n f(z_n) = \int_{C_0} dz
  \mbox{tanh} [\beta z/2 ] f(z)$,
where $f[z_n\equiv i(2n+1)\pi\beta^{-1}]$ is the logarithmic function in
Eq. \ref{MatsuAct}, $C_0$ is shown in Fig. \ref{Fig:Matsu} and the
proper $f(z)$ that gives the correct thermodynamics is
$f(z) = \ln [m^2+\mbox{\boldmath{$p$}}^2 - (z+\mu)^2]-\ln [
    m^2+\mbox{\boldmath{$p$}}^2 - z^2]$.
Since $f(z)$ approaches to zero fast enough in the $|z|\to\infty$
limit and is analytic in the complex $z$ plane excluding the real
axis, the integration over $z$ along contour $C_0$ is equivalent to
the sum of integration over contours $C_+$ and the negative of
$C_-$, which gives $\Omega= U-TS-\mu N=-PV$ expected from thermodynamics.

    The reason why the FMS path integration evaluation for the r.h.s.
of Eq. \ref{PartFun1} using a 4 component fermion fields fails to give
the correct thermodynamics at zero temperature
can be understood by a comparison of Figs.
\ref{Fig:Cont1} and \ref{Fig:Matsu}. The  causal structure of
the FMS path integration restricts the $p_0$ integration for
the partition function to be within the class of contours shown in Fig.
\ref{Fig:Cont1}. It lacks the contour pair $C_+$ and
$C_-$ implied by the
boundary condition Eq. \ref{Fermion_BC} shown in Fig. \ref{Fig:Matsu},
which are necessary components to given the correct thermodynamics.
With the 8 component ``real'' spinor $\Psi$ to represent the
fermion fields, the effects of the $C_-$ are provided by the
lower 4 component $\psi_2$ of $\Psi$.

We are  ready to develop a real-time field theory
at finite density and temperature using the 8 component ``real''
representation for the fermion fields.
The boundary condition Eq. \ref{Fermion_BC} is written in an
equivalent form. For the fermions interested in this study, it can be
expressed as
$\widehat\Psi(t) = -e^{\beta\widehat K}\widehat\Psi(t) e^{-\beta\widehat K}$.
It is equivalent to Eq. \ref{Fermion_BC} due to the fact
that the conserved $\widehat N$ commutes with the total Hamiltonian $\widehat
H$, which allows the factorization of the action of $\beta\mu\widehat N$
and $\beta\widehat H$ in the exponential of $\beta\widehat K$ with the
$\exp(\beta\mu)$ factor\footnote{It should be exp($\beta\mu O_3$) in
the new representation.} in Eq.  \ref{Fermion_BC} the result of the
action of $\exp(\beta\mu\widehat N)$ and $\exp(-\beta\mu\widehat N)$ on $\psi$
from left and right. To handle non-perturbative symmetry breaking
cases, the commutativity of $\widehat H$ and $\widehat N$ shall not be imposed
at this level of development but rather at later stages as dynamical
constraints.

The real-time thermal
field dynamics in the 8 component ``real'' representation for fermion
fields can be constructed along the same line as that of Refs.
\cite{LandsWeer}, \cite{Kapusta}, \cite{RTTFD1}--\cite{RTTFD3}.
Compared to those developments,
the only formal differences of the present approach are 1) the
Kubo--Martin--Schwinger (KMS) boundary condition for the contour
propagator do not contain the $\exp(\beta\mu)$ factor in the present
approach, 2) the effects of $\mu$ is hidden in the energy variable
within the propagator (namely, the time evolution is generated by
$\widehat K$ not $\widehat H$) here and 3) the symmetry factor for a Feynman
diagram in a perturbation expansion is
different from the 4 component representation for the fermion fields
due to Eq. \ref{realcond}, which also allows the Wick
contraction between two $\Psi$s \cite{Paps}.
The matrix form of contour
propagator\footnote{The returning time contour parallels the real time
  axis is chosen to be located half way between $0$ and $-i\beta$.} for a
fermion is
\begin{eqnarray}
  {\cal S}(p) & = & M \left ( \begin{array}{cc}
                   S^{+}&0\\0& S^{-}\end{array} \right )
                  M,
\label{Mat-Prop}
\end{eqnarray}
where the retarded and advanced propagators $S^+$ and $S^-$ are
$S^\pm(p) = i/( \rlap\slash p +
    \gamma^0 \mu O_3 - m \pm i \gamma^0\epsilon )$ and
$M=\mbox{sgn}(p_0) \sqrt{1-n(p_0)}-i\sigma_2 \sqrt{n(p_0)}$ with
$n(e) = 1/[exp(\beta e)+1]$ and $\sigma_2$ the second Pauli matrix.
The poles of the above propagator at zero temperature are
located above the contour $C_{FMS}$ if they are negative and below
it otherwise. It can be
compared to the conventional real-time
approach with the correct causal structure
\cite{Shuyak,Lutz} at zero temperature, in which the poles of the
fermion propagator are
located above the $p_0$ integration contour if they are on the left
of $\mu$ or else below it.
The real-time partition function obtained here for
the system is identical to the Matsubara method
since it has a better large $p_0$ behavior at finite density.

 Let us turn  to the study of local bilinear operators constructed from
two fermion fields of the form $\widehat O = \bar\psi(x)\Gamma\psi(x)$
in a finite
density situation where $\Gamma$ is certain matrix acting on $\psi$.
A product of two field operators is in general singular and non-unique.
A definition of
such a potentially divergent product is
$
   \widehat O = \lim_{\delta_\mu\to 0} \bar\psi(x+\delta)\Gamma\psi(x),
$
 where $\delta_\mu$ is a 4-vector with, e.g., $\delta_0<0$.
The ground state (vacuum) expectation value of $\widehat O$ can thus be
computed from the fermion propagator ${\cal S}$ by closing the
integration contour on real $p_0$ axis in the lower half of the complex $p_0$
plane.
Since there is no poles and cuts off the real $p_0$ axis except the
ones on the imaginary $p_0$ axis due to the thermal factor, which
should be excluded (see Fig. \ref{Fig:Matsu}), it can be deformed to $C_R$ to
include the poles and cuts of the retarded propagator $S^+$:
\begin{eqnarray}
   <\widehat O> &=& - \mbox{tr} \int {d^3p\over (2\pi)^3}\int_{C_R}
                         {dp_0\over 2\pi} {\cal S}^{11}_F(p)\Gamma,
\label{avO}
\end{eqnarray}
with ``tr'' denoting the trace over internal indices of the fermion
fields. $C_R$ reduces to $C_+$ as $\beta\to 0$.
For a free system, there are only poles in ${\cal S}^{11}$ so
that Eq. \ref{avO} can be expressed as
\begin{eqnarray}
  < \widehat O > &=& i\int {d^3p\over (2\pi)^3}\sum_k [1-n(p^0_k)]
                   \mbox{tr}\mbox{Res}
                    S^{+}(p_k)\Gamma,
\label{avO1}
\end{eqnarray}
where the sum is over  all poles $p^0_k$ of $S^{+}(p)$ and
``Res'' denotes the corresponding residue.
The difference between the 8 component theory
here and the 4 component one also manifests here. For example, the
conserved fermion number density $\bar n$ corresponding to current $j^\mu =
{1\over 2} \bar\Psi \gamma^\mu O_3 \Psi$ is
$
       \bar n  =  {1\over\pi^2}\int^\infty_0 d\mbox{\boldmath $p$}
                  \mbox{\boldmath $p$}^2
                 \left [f^{(-)}_p - f^{(+)}_p\right ]
$
with $f^{(\pm)}_p = 1/[e^{\beta(E_p\pm \mu)}+1]$.
It is same as the one in elementary statistical mechanics.
On the other hand, $\bar n$ computed in the conventional approach
using such a point split definition of fermion number density is
divergent and different from each other for $\delta_0>0$ and
$\delta_0<0$, which means that it is not even consistent with relativity
it starts with 
for a space like $\delta_\mu$ whose time component can has different
sign in different frames. It can be made finite and unique only 
after an arbitrary subtraction.

 The particle content of the 8 component theory
can be found by studying pole structure of the
fermion propagator together with the consideration of the FMS causal
condition. The time dependence of the propagator at zero temperature is
${\cal S}(t,\mbox{\boldmath $p$}) = \int_{C_{FMS}} {dp_0\over 2\pi} 
e^{-ip_0 t} {\cal S}(p_0,\mbox{\boldmath $p$})$,
which, in case of $t>0$, can be evaluated on the contour $C_+$ with a result
\begin{eqnarray}
   {\cal S}(t,\mbox{\boldmath $p$}) 
               &=& \theta(E_p-\mu) \Lambda^{1+}_p e^{-i(E_p-\mu)t} +
               \theta(\mu-E_p) \Lambda^{2-}_p  e^{-i(\mu-E_p)t} + 
               \Lambda^{2+}_p e^{-i(E_p+\mu)t}  
\label{PropTgt0}
\end{eqnarray}
and, if $t<0$, can be evaluated on the contour $C_-$ to obtain
\begin{eqnarray}
   {\cal S}(t,\mbox{\boldmath $p$}) 
               &=& \Lambda^{1-}_p  e^{i(E_p+\mu)t} +
               \theta(\mu-E_p) \Lambda^{1+}_p e^{i(\mu-E_p)t} + 
               \theta(E_p-\mu) \Lambda^{2-}_p e^{i(E_p-\mu)t}.    
\label{PropTlt0}
\end{eqnarray}
Here $\Lambda_p^{r\pm} = P_r(\pm\gamma^0 E_p - \mbox{\boldmath
  $\gamma$}\cdot \mbox{\boldmath $p$} + m)/2E_p $, $P_1=(1+O_3)/2$
and $P_2=(1-O_3)/2$ are projection operators.

The FMS causal structure (or boundary condition) 
in the present theory can be simply putted as: 1)
excitations with $p_0> 0$ that propagate 
{\em forward in time} correspond to particles or antiholes and 2) those with 
$p_0< 0$ that propagate {\em backward in time} 
correspond to antiparticles or holes. The quantization of the 8 
component fermionic field then follows naturally. 
In case of zero density ($\mu=0$), $\widehat \Psi(x)$ can be written as
\begin{eqnarray}
    \widehat \Psi(x) &=& \sum_{rps} {1\over 2 E_p}
       \left [ U_{rps} e^{-ip\cdot x} \hat b_{rps} +
           V_{rps} e^{ip\cdot x} \hat {d^\dagger}_{rps}\right ].
\label{PsiOp}
\end{eqnarray}
Here $s$ is the spin index, $r=1,2$,
$U_{rps}$ and $V_{rps}$ are 8 component spinors that satisfy
$(\rlap\slash p - m)U_{rps}=0$, $O_3U_{rps}=(3-2r)U_{rps}$,
$(\rlap\slash p + m)V_{rps}=0$, $O_3V_{rps}=(3-2r)V_{rps}$
and the non vanishing anticommutators between
$\hat b_{rps}$ and $\hat d_{rps}$ are
$   \left \{\hat b_{rps},\hat b^\dagger_{r'p's'} \right \} =
        2E_p\delta_{pp'}\delta_{ss'}\delta_{rr'}$,
$\left \{\hat d_{rps},\hat d^\dagger_{r'p's'} \right \} =
        2E_p\delta_{pp'}\delta_{ss'}\delta_{rr'}$,
which lead to a canonical quantization of $\widehat \Psi$. The particle states 
are obtained from the vacuum state $\mathclose{|\,0\rangle}$, which is annihilated
by the $\hat b$s and $\hat d$s above, by the action of $\widehat \Psi$
and $\widehat {\bar\Psi}$ on it. Albeit proper counting has already been 
given to the intermediate states during its evolution 
\cite{Paps}, these particle states are not all detectable
since the reality condition Eq. \ref{realcond} has to be 
imposed on the detector. The operator reality condition for the 
fermion field $\widehat \Psi$ are given by the mirror reflection
operator $\widehat {\cal M}$, which is defined by $\widehat {\cal M}
\widehat {{\bar \Psi}^T} \widehat {\cal M} \equiv {\cal M} \widehat
{{\bar\Phi}^T}$ with $\widehat M^2 =\widehat I$,
where $\widehat {\bar\Phi}$ is obtained from $\widehat\Psi$ by
the following replacement: $\widehat {\cal M}\hat b_{rps}\widehat {
\cal M}=\hat d_{rps}$.
The operator reality condition corresponding to Eq. \ref{realcond}
is then $\widehat \Psi = {\cal M} \widehat {{\bar \Phi}^T}$.
The vacuum state satisfies $\widehat P_{phys} \mathclose{|\,0\rangle}
=\mathclose{|\,0\rangle}$. 
So the physical states satisfying the operator reality condition
are those ones symmetric in $r=1$ and $r=2$ 
projected out by $\widehat P_{phys} = (\widehat I + \widehat {\cal M})/2$.
In finite density case, the annihilation and
creation operators for the holes (in the Fermi sea) has to be introduced, which
is straightforward.

Compared to the
4 component theory, which has only one fermionic excitation excitation mode with an effective
energy $\epsilon_{-}=E_p-\mu$ ($\mu>0$) for particle,
which is the one with $r=1$ in Eq. \ref{PropTgt0},
Eq. \ref{PropTgt0} has a new fermionic excitation mode
with $r=2$ and an effective energy $\epsilon_{+}=E_p+\mu$ possessing a {
common 3 momentum range, spin structure and opposite charge} against the
$r=1$ one. This excitation mode
is much larger in energy than the $r=1$ one ($\epsilon_{-}$) in a
non-relativistic system like the electron gas, it can be comparable in
relativistic systems like the quark gluon plasma where the bare mass
of the fermion is very small relative to its momentum. In addition,
there is another particle mode with $r=2$ and effective energy
$\mu-E_p$. Its effects will be further discussed in the following.
First, let us mention that the $r=1$ and $r=2$ parts
in a free theory decouple and give, separately, identical results for
observables due to the charge conjugation symmetry. They belong to two non
communicating opposite charged mirror universes.
Therefore the differences between the 8 component theory
and the 4 component one can not show themselves if the $r=1$ and $r=2$
parts are not produced in coherent states of the form
\begin{eqnarray}
      \mathclose{|\,\phi\rangle} &=& \cos\phi
         \mathclose{|\,1\rangle} + e^{i\delta} \sin\phi
         \mathclose{|\,2\rangle}
\label{quant-state}
\end{eqnarray}
with $\phi\ne 0 $ and $\delta$ arbitrary, which does not separat apart 
during its evolution in time.  The following is a scenario
in which quantum interference effects of the 8 component theory that
are not present in the 4 component one can be observed for interacting
theories.

The possibility of the existence of vacuum phases in a relativistic
massless fermion system induced by a condensation of fermion pairs and
antifermion pairs and their possible
relevance to physical hadronic system are studied in Refs. \cite{Paps,report}.
Suppose that the $\beta$ and $\omega$
phases discussed in Refs. \cite{Paps,report} that mix the $r=1$ and 
$r=2$ mirror universes are produced in, e.g. a
heavy ion collision, or an astronomical object,
then a quantum state of the form given by Eq.
\ref{quant-state} with a fixed $\delta$ and $\phi\ne 0$ can be
produced in the collision region \cite{Paps}. Such a state contains $r=1$ and $r=2$ parts
which have the same 3 momentum $\mbox{\boldmath $p$}$. After this 
state leaves the collision region where the density is
effectively zero to propagate to the detector, they remains in a mixed
state of the form given by Eq. \ref{quant-state} with $r=1$ and
$r=2$ parts having the same 3 momentum, which guarantees that these two
parts can propagate together without been spatially separated
on their way to the detector. To distinguish
these two components of the particle in the detector, one can generate
a non zero electric potential $A^0$ by, say, putting the detector into a
hollow charged conductor. Inside this $A^0\ne 0$ region, 
the $r=1$ and $r=2$ pieces begin to have a different time
dependence $\exp[-i(E_p\mp q A^0) t]$ with $q$ the charge of the particle, 
which can generates detectable beats in the counting rate. 
To show this more explicitly, let us
consider the counting rate of finding a particle in a space volume
$\Delta\Omega$;   $P_{\Delta\Omega}=\int_{\Delta\Omega} d^3x \left |
    \mathopen{\langle x\,|}
    \mathclose{\,{\mbox{\boldmath $p$},\phi}\rangle} \right |^2$ is
\begin{eqnarray}
&&       \int_{\Delta\Omega}d^3x \left [\cos^2\phi \left 
          |\mathopen{\langle x\,|}
          \mathclose{\,\mbox{\boldmath $p$},1\rangle} \right |^2
         + \sin^2\phi \left |\mathopen{\langle x\,|}
           \mathclose{\,\mbox{\boldmath $p$},2\rangle} \right |^2
         \right . \nonumber\\ && \left .
\phantom{\mathopen{\cos^2} \mathclose{0}} 
             + \sin\phi \cos\phi \left ( \mathopen{\langle x\,|}
             \mathclose{\,\mbox{\boldmath
             $p$},1\rangle} 
             \mathopen{\langle x\,|}
             \mathclose{\,\mbox{\boldmath $p$},2\rangle}^* e^{-i\delta} 
             + c.c \right
             )\right ],
\end{eqnarray}
where $\mbox{\boldmath $p$}$ is the 3 momentum of the excitation,
$c.c.$ denotes complex conjugation and the single particle physical 
state at $x$ satisfies $\mathopen{\langle x\,|}\widehat P_{phys} 
= \mathopen{\langle x\,|}$.
It is ready to see that the third
term depends on time with an universal beat frequency $2qA^0$ for all
directly pruduced particles with the same $q$ when $\phi\ne 0$. This
effect is absent if the conventional 4 component theory is used. 

The other excitation mode with $r=2$ in Eq. \ref{PropTgt0} that corresponds to
particle has a different spin structure and momentum range from the 
$r=1$, $\epsilon_{-}$ excitation mode. This excitation mode is not expected to has
any interference effects with the $r=1$, $\epsilon_{-}$ excitation mode. 
Their effects on physical observables are addative on the
probability level instead of the amplitude level. So they simply
reproduce the effects of those in the 4 component theory due to charge
conjugation symmetry of the system.

 In summary, we developed an 8 component theory for the relativistic finite
density systems that has better short distance behavior than the corresponding
4 component one. The introduction of a mirror universe is found to be 
necessary. The link between the possible mirror universe and ours can be 
established and remains to be searched for in the real world phenomena of 
heavy ion collisions, mechanism for the CP violation, baryogenesis in the
early Universe, the origin of dark matter, etc.

\section*{Figures}
\begin{figure}
\caption{\label{Fig:Cont1} The set of contours belonging to the same
  FMS class. Contour $C_{FMS}$ is
  the original FMS contour in the Minkowski space. Contour $C_E$ is
  the Euclidean contour. Contour $C_+$ is the quasiparticle contour.
   $\pm i\pi$ denote the imaginary part of the integrand along the
   edges of its cuts (thick lines) on the physical $p_0$ plane.}
\end{figure}
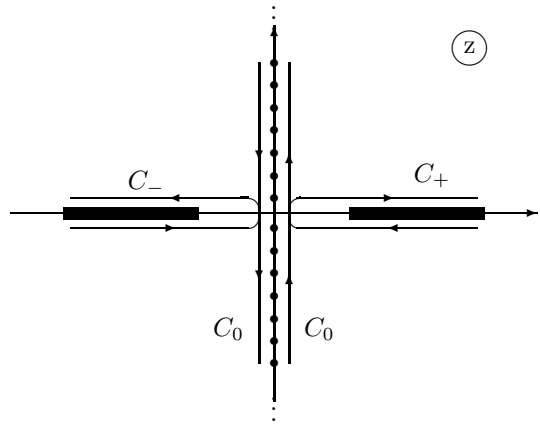
\begin{figure}[h]
\caption{\label{Fig:Matsu} The set of contours that give the
  thermodynamics. Contour $C_0$ encloses the Matsubara poles of the
  integrand. Contours $C_+$ and $C_-$ are the ones needed in a real
  time formulation of the thermal field theory in the 4 component
  representation for fermion fields.}
\end{figure}
\newpage

\begin{figure}[h]
\unitlength=0.80mm
\linethickness{0.5pt}
\begin{picture}(120.00,113.00)(-27,10)
\put(40.00,80.00){\vector(1,0){80.00}}
\put(124.00,75.00){\makebox(0,0)[cc]{$Rep_0$}}
\put(86.00,111.00){\makebox(0,0)[cc]{$Imp_0$}}
\put(90.00,78.95){\rule{24.00\unitlength}{2.10\unitlength}}
\put(46.00,78.95){\rule{24.00\unitlength}{2.10\unitlength}}
\put(90.00,80.00){\circle*{1.50}}
\put(70.00,80.00){\circle*{1.50}}
\put(42.00,78.00){\line(1,0){36.00}}
\put(78.00,79.00){\oval(4.00,2.00)[rb]}
\put(82.50,81.00){\oval(5.00,2.00)[lt]}
\put(82.00,82.00){\line(1,0){36.00}}
\put(96.00,82.00){\vector(1,0){4.00}}
\put(57.00,78.00){\vector(1,0){5.00}}
\put(80.00,48.00){\vector(0,1){65.00}}
\put(80.00,62.00){\vector(0,1){2.00}}
\put(80.00,71.00){\vector(0,1){2.00}}
\put(80.00,90.00){\vector(0,1){2.00}}
\put(80.00,98.00){\vector(0,1){3.00}}
\put(83.00,83.00){\line(1,0){32.00}}
\put(83.00,77.00){\line(1,0){32.00}}
\put(83.50,80.00){\oval(7.00,6.00)[l]}
\put(90.00,83.00){\vector(1,0){2.00}}
\put(104.00,83.00){\vector(1,0){2.00}}
\put(106.00,77.00){\vector(-1,0){2.00}}
\put(95.00,77.00){\vector(-1,0){2.00}}
\put(64.00,74.00){\makebox(0,0)[cc]{$C_{FMS}$}}
\put(75.00,93.00){\makebox(0,0)[cc]{$C_E$}}
\put(97.00,74.00){\makebox(0,0)[cc]{$C_+$}}
\put(110.00,85.00){\makebox(0,0)[cc]{$-i\pi$}}
\put(110.00,74.00){\makebox(0,0)[cc]{$+i\pi$}}
\put(48.00,74.00){\makebox(0,0)[cc]{$-i\pi$}}
\put(48.00,85.00){\makebox(0,0)[cc]{$+i\pi$}}
\end{picture}
\end{figure}
\begin{center} Figure 1 \end{center}
\newpage
\begin{figure}[h]
 \unitlength=1.00mm \linethickness{0.5pt}
\begin{picture}(110.00,107.00)(-5,10)
\put(45.00,80.00){\vector(1,0){70.00}}
\put(80.00,55.00){\vector(0,1){50.00}}
\put(80.00,60.00){\circle*{1.10}}
\put(80.00,63.00){\circle*{1.10}}
\put(80.00,66.00){\circle*{1.10}}
\put(80.00,69.00){\circle*{1.10}}
\put(80.00,72.00){\circle*{1.10}}
\put(80.00,75.00){\circle*{1.10}}
\put(80.00,78.00){\circle*{1.10}}
\put(80.00,82.00){\circle*{1.10}}
\put(80.00,85.00){\circle*{1.10}}
\put(80.00,88.00){\circle*{1.10}}
\put(80.00,91.00){\circle*{1.10}}
\put(80.00,94.00){\circle*{1.10}}
\put(80.00,97.00){\circle*{1.10}}
\put(80.00,100.00){\circle*{1.10}}
\put(90.00,79.20){\rule{18.00\unitlength}{1.60\unitlength}}
\put(52.00,79.20){\rule{18.00\unitlength}{1.60\unitlength}}
\put(82.00,60.00){\line(0,1){40.00}}
\put(82.00,70.00){\vector(0,1){2.00}}
\put(82.00,85.00){\vector(0,1){3.00}}
\put(78.00,100.00){\line(0,-1){40.00}}
\put(78.00,89.00){\vector(0,-1){2.00}}
\put(78.00,74.00){\vector(0,-1){3.00}}
\put(75.50,80.00){\oval(5.00,4.00)[r]}
\put(83.50,80.00){\oval(3.00,4.00)[l]}
\put(69.00,82.00){\vector(-1,0){3.00}}
\put(64.00,78.00){\vector(1,0){3.00}}
\put(106.00,102.00){\circle{4.47}}
\put(106.00,102.00){\makebox(0,0)[cc]{z}}
\put(101.00,85.00){\makebox(0,0)[cc]{$C_+$}}
\put(63.00,84.00){\makebox(0,0)[cc]{$C_-$}}
\put(80.00,55.00){\makebox(0,0)[cc]{$\vdots$}}
\put(80.00,107.00){\makebox(0,0)[cc]{$\vdots$}}
\put(84.00,66.00){\makebox(0,0)[lt]{$C_0$}}
\put(76.00,66.00){\makebox(0,0)[rt]{$C_0$}}
\put(53.00,82.00){\line(1,0){23.00}}
\put(53.00,78.00){\line(1,0){23.00}}
\put(94.00,82.00){\vector(1,0){2.00}}
\put(97.00,78.00){\vector(-1,0){2.00}}
\put(83.00,82.00){\line(1,0){24.00}}
\put(83.00,78.00){\line(1,0){24.00}}
\end{picture}
\end{figure}
\begin{center} Figure 2 \end{center}

\end{document}